# Mining Evidence about Your Symptoms: Mitigating Availability Bias in Online Self-Diagnosis

Junti Zhang
National University of Singapore
Singapore
juntizhang@u.nus.edu

Zicheng Zhu
National University of Singapore
Singapore
zicheng@u.nus.edu

Jingshu Li
National University of Singapore
Singapore
jingshu@u.nus.edu

Yi-Chieh Lee
National University of Singapore
Singapore
yclee@nus.edu.sg

## ABSTRACT

People frequently exposed to health information on social media tend to overestimate their symptoms during online self-diagnosis due to availability bias. This may lead to incorrect self-medication and place additional burdens on healthcare providers to correct patients' misconceptions. In this work, we conducted two mixed-method studies to identify design goals for mitigating availability bias in online self-diagnosis. We investigated factors that distort self-assessment of symptoms after exposure to social media. We found that availability bias is pronounced when social media content resonated with individuals, making them disregard their own evidences. To address this, we developed and evaluated three chatbot-based symptom checkers designed to foster evidence-based self-reflection for bias mitigation given their potential to encourage thoughtful responses. Results showed that chatbot-based symptom checkers with cognitive intervention strategies mitigated the impact of availability bias in online self-diagnosis.

## 1 INTRODUCTION

Access to online health information has become increasingly prevalent among the public. Social media users, through passive exposure to health-related posts and active seeking of health information, are becoming more inclined to make judgments and decisions about their own health status [53, 96]. However, compared to other health information sources such as from friends or family, exposure to health-related information on social media has the potential to amplify public concern and lead to over-diagnosis of diseases [1, 6], especially due to the continuous exposure to similar content driven by recommendation algorithms [87]. In particular, emotionally charged or frequently seen information about specific health condition seems more familiar and is likely to be overestimated [43]. This phenomenon is identified as *availability bias*, i.e., overestimating event likelihood due to ease of recall [82]. It is caused by people's reliance on heuristic strategy rather than analytical thinking [83], as it costs less time and effort to make a judgment [51]. In short, exposure to health information on social media can trigger availability bias and potentially lead to inaccurate self-assessment.

With the growing popularity of online symptom checkers (OSCs) — which now include questionnaires, live consultations, and chatbot-driven platforms [91] — people increasingly self-diagnose before seeking medical advice [26]. However, this trend presents challenges for the healthcare industry [22] as patients with assumed diagnoses may resort to incorrect self-medication without guidance. Meanwhile, facing patients with a presumed diagnosis could hinder the patient-doctor relationship and complicate the ability of physicians to perform their duties effectively, as they often need to contradict the beliefs patients have obtained online [8, 39, 46]. Given the potential role of availability bias in shaping inaccurate self-diagnosis, it is crucial to study the impact of social media exposure to inform the design of interventions that can mitigate these effects.

Prior studies delved into the strategies of mitigating such bias [15, 48, 49], aiming to enhance diagnostic accuracy by promoting self-reflection, which requires more mental effort and evidence-based reasoning. Despite these advances, little is known about how availability bias induced by social media exposure affects individuals without professional medical knowledge. Moreover, a significant research gap persists in mitigating availability bias during self-assessment with OSC, assuming that individuals have a suitable perception of their health conditions after exposure to social media. To explore the impact of social media exposure on users and inform the design of effective bias mitigation strategies, we aim to address the first research question:

**RQ1:** How does the health-related information on social media trigger availability bias in online self-diagnosis?

To address RQ1, we conducted a between-subject experiment (N=104) to investigate how availability bias could be triggered by exposure to health-related social media posts during a self-assessment task for adult ADHD. We compared the impact of neutral and exaggerated content to a controlled condition, since they represent the most common and highly impactful types of health information on social media [57, 85]. We found that exposure to neutral content induced the most significant availability bias and led participants to overestimate their symptom levels. In particular, the relevance caused them to disregard their own evidence during self-assessment. To address this, fostering evidence-based self-reflection during symptom self-assessment could be a potential strategy to reduce availability bias.

Chatbots have recently been shown to have the ability to effectively foster re-examination on the user's beliefs [17, 77], and guide one to reflect on evidence or different perspectives [65] due to its conversational and interactive nature. Users often invest more effort and engagement in providing additional evidence to support their perspectives through dialogues [13], making chatbots a growing focus of research in the HCI community and increasingly integrated into real-world social media platforms [2, 76]. Therefore, we proposed three design strategies in the form of the questioning-and-answering frameworks for bias mitigation with chatbot-based symptom checkers (CSCs) that promote user's self-reflection. More specifically, we seek to answer the second research question:

**RQ2:** To what extents do the CSCs designed to promote self-reflection mitigate availability bias in online self-diagnosis?

To address RQ2, we seek to prompt self-reflection with three designs of symptom checkers: CSC, CSC with Evidence Reflection, CSC with Counterfactual Thinking. In the second study (N=100), we explored whether and how these CSCs with cognitive intervention strategies mitigate availability bias during self-diagnosis through a mixed-method experiment. Our results showed that the CSCs with cognitive interventions were effective in mitigating availability bias by guiding the user to engage in an evidence-based reflective thinking about their health conditions.

In summary, we contribute the following through this work:

- We identify potential causes behind overestimation of symptom due to social media exposure as two main reasons, thus underscoring how availability bias may affect symptom self-assessment and enriches our understanding of users' susceptibility to this bias under social media exposure.
- We propose integrating cognitive intervention strategies into CSCs and empirically demonstrate their effectiveness in mitigating availability bias in online self-diagnosis.
- Our findings provide new implications into designing and using chatbots where evidence-based self-reflection is needed.

## 2 RELATED WORK

### 2.1 Health Information on Social Media and Self-Diagnosis

Over the last two decades, the trend of using the social media platforms for health information seeking is on the rise [96]. Social media effectively allows individuals, including patients and their families, to gather information, seek assistance, support others, obtain forum support, and share personal experiences. As indicated by these studies [37, 38], obtaining health information from social media has become a key factor in driving health-related behavioral changes.

Health information on social media shapes user's perception and belief, therefore plays as an important role in guiding individuals toward seeking medical advice and performing self-diagnosis. Self-diagnosis refers to the process by which individuals attempt to identify their medical conditions using tools like OSCs without the intervention of healthcare professionals [22]. This trend has become a challenge for the healthcare industry [54, 89], as the lack of medical training can result in wrong ways of self-medicating and harming doctor-patient relationship [21, 29, 45].

Recent studies have further emphasized this challenge by proving that misleading content tends to thrive in all platforms through users' sharing of unvalidated information [47]. The most common misleading health information on social media is exaggeration [3, 84]. It involves the addition or alteration of the severity or prevalence of health information through the use of superlatives or data [81]. Misleading health information even spread more rapidly and exerts a greater impact on the public than comparable authentic content on social media [85], and thus has the potential to trigger more cognitive bias [57].

Prior studies have primarily focused on examining the types and prevalence of misleading information within social media communities, as well as the accuracy of self-diagnoses among non-health professionals. However, few researchers have considered how individuals perceive health-related social media content and how these perceptions influence their self-assessment on symptom levels. Based on existing theories, we can expect that participants exposed to neutral social media content will experience distortions in their self-assessment, with those exposed to exaggerated content likely experiencing even greater distortions. Specifically, we pose the following hypothesis:

**H1:** Social media users' exposure to neutral health information about specific diseases will skew their self-assessment (H1.a). Additionally, users expose to exaggerated health information will be affected to a greater extent than those who encounter neutral information (H1.b).

### 2.2 Availability Bias in Healthcare

Availability bias occurs when individuals overestimate the likelihood of events based on their recall ability, amplified by recent social media exposure [82]. Several mechanisms explain how exposure influences perceptions. First, message learning occurs when repeated exposure enhances recall [94]. Second, repeated exposure increases importance attribution to health conditions during self-assessment [9]. Third, high levels of exposure create implicit social norms, especially when information is amplified by recommendation algorithms [30]. Human susceptibility to availability bias could be explained by dual-system theory: System 1, intuitive and heuristic-based; and System 2, deliberate and analytical [33, 86]. People often use heuristics to save time and effort [83], which is usually effective. However, heuristics can lead to errors and cognitive biases by ignoring facts and data [50]. Studies in the field of health of availability bias focus primarily on healthcare professionals [43], showing its impact on clinical judgments. This requires physician training to address the bias during assessments [97]. Susceptibility to biased online information also relates to age [73], educational background [70], and health literacy, which is the ability to understand medical information [4, 52].

Recent advances in cognitive psychology and clinical decision support systems have explored bias mitigation techniques such as checklists [20], cognitive forcing strategies [14], diagnostic time-outs [79], and slow-downs [10]. Reilly et al. [67] showed that long-term education on cognitive biases and diagnostic error significantly improved diagnostic accuracy. These approaches target availability biases by promoting self-reflection, requiring more mental effort and evidence-based reasoning, foundational for effective interventions. However, a significant research gap remains in addressing availability bias in non-medical professionals. Gocko et al. [25] examined the Internet's role and cognitive biases in chronic Lyme disease controversies but had not explored why patients are susceptible to social media influence or suggest strategies to mitigate this bias. This paper hypothesizes that social media-induced availability bias affects users' diagnostic outcomes. Specifically, with a focus particularly on self-diagnosis, this work posits that:

**H2:** People will overestimate their symptom levels of a disorder under social media exposure. The exposures include neutral (H2.a) and exaggerated (H2.b) content.

## 2.3 Chatbots for Promoting Reflective Thinking and Mitigation of Cognitive Bias

Chatbots are designed to interact with people using natural language and can automatically provide information upon request [93]. Chatbots have shown the ability to encourage higher-quality information elicitation [90] and self-disclosure [40] from online users. This can be explained by the Computers Are Social Actors (CASA) paradigm that people mindlessly apply the social norms and expectations of human relationships when interacting with computer agents like chatbots [60]. The HCI community has recently concentrated on utilizing chatbot to facilitate self-reflection. For example, Mukherjee et al. [58] developed a fine-tuned language model to enhance the creation of impact statements by promoting self-reflection. An AI-based questioning framework proposed by Danry et al. [17] has been proven to actively engage users' thinking and support their reasoning process. Furthermore, HCI researchers are increasingly using chatbots to expose users to different perspectives and encourage them to reconsider their initial attitudes and positions [19, 77]. These studies collectively suggest that chatbots could be promising tools to assist users in self-reflection.

In healthcare context, chatbots have been used to solve the survey fatigue problem [61] that impact the quality of information collected from patients [41, 42]. This occurs because users often invest considerable effort and engagement in establishing a common ground with chatbots through dialogues [13], thus made them as a superior alternative to static web-based surveys [36]. With these benefits, the integration of online symptom checkers and chatbots has become a helpful diagnostic tool through interactive dialogue [66], aiding self-triage, providing potential diagnoses, and offering medical advice [56, 72]. Recent studies explored CSCs [75, 92] highlighted emotional support, explainability, and efficiency as key factors for users. Another study added post-hoc explanations to a CSC, helping users understand the rationale behind certain questions [80]. Explanations improved perceived diagnostic quality and trust, but increased cognitive overload affected user experience.

Despite the attention to CSC user experience, it remains unclear how chatbot design can address cognitive biases in non-health professionals. Our research contributes by designing a CSC that promotes self-reflection during diagnosis. We investigate availability bias in self-diagnosis and provide insights into the effect of CSCs with cognitive intervention strategies.

## 3 STUDY 1 METHOD: EFFECTS OF AVAILABILITY BIAS INDUCED BY SOCIAL MEDIA EXPOSURE

To address **RQ1**, we conducted a between-subject experiment, where participants were exposed to three types of health experiences shared on the simulated social media. We compared the influences of social media exposure on self-assessment, diagnostic outcomes, perceived information reliability, and their perceptions through qualitative follow-up across groups.

### 3.1 Conditions

We investigated three conditions that varied in whether and how the shared health experiences are exposed to the participants. Given that social media, driven by recommendation systems, tends to push similar content that aligns with a user's prior searches [12], the neutral condition allowed us to simulate users' exposure to factual information under such circumstances. The exaggerated condition simulates the heightened emotional and cognitive impact of sensationalized information, which has been shown to distort users' perceptions of health risks and conditions [57]. This design enables a systematic investigation of how varying levels of content influence cognitive mechanisms like availability bias, providing a comprehensive understanding of these dynamics within the context of social media use. By comparing these conditions, we could assess how different levels of content bias affect availability bias, thereby shedding light on the underlying mechanisms.

(1) *Control:* Participants were shown social media posts with random, non-selective health information about common disorders. The control group was designed to establish a baseline for understanding the natural self-assessment process without selective exposure under recommendation algorithms.
(2) *Neutral:* Participants were exposed to health experiences shared by other social media users who had experienced the symptoms of a same disease. The neutral contents were selected to be factual and unbiased, providing accurate and diverse views without misinformation. This condition simulates personalized information push after a social media user made several searches for certain health condition under recommendation algorithm [68].
(3) *Exaggerated:* Given that health-related misleading information on social media exerts a greater impact and leads to poor health decisions[63, 85], we selected exaggeration in health-related content [95] as it is the most frequently presented feature of health misinformation on social media [44]. Participants in this group were shown posts contain exaggerated facts, which the information provided exaggerated the severity and the prevalence of a same disorder.

## 3.2 Experimental Materials

We selected adult ADHD to be diagnosed in this study according to these criteria: (1) possible to obtain preliminary diagnoses through self-reported data from patients [5]; (2) feasible to involve general participants without the need for screening subjects with certain symptoms, i.e., symptoms are common and ambiguous among the public [62]; and (3) contains no sensitive medical information. Adults with ADHD show symptoms like attention deficit, impulsiveness, and forgetfulness, which can be mistaken for non-pathological behaviors [24]. This allowed us to include a broader population without need for specific recruitment of symptoms.

*3.2.1 Implementation.* We developed a web-based platform designed to replicate the experience of navigating social media environments with HTML and JavaScript. The interface of the platform was specifically simulated from a version of the social media user interface. This design aimed to leverage users' existing familiarity with widely recognized social media layouts to ensure immediate ease of use. The simulation allowed for controlled exposure to specific types of content according to each experimental group. The screen shots of the experimental interface are shown in Figure 1. Considering the average length of social media posts, participants in all groups were required to read through 5 posts on the simulated platform, as this number ensures a sufficient amount of information is presented without inducing cognitive fatigue, consistent with prior CHI studies on social media disclosures that have adopted a similar approach [69]. The full simulated posts for all groups can be found in Appendix.

Additionally, participants were not obliged to interact with the posts by liking them or leaving comments, though they were able to do so if they wanted.

*3.2.2 Collection of social media posts.* To construct the posts for the two test groups, three researchers with interdisciplinary expertise in HCI and healthcare collected authentic user contributions from communities dedicated to ADHD on Reddit[1], Twitter[2], and Facebook[3]. We selected these platforms because they are widely recognized within the HCI community as popular, open social media spaces for accessing and sharing health-related information [35]. To query for posts related to adult ADHD and its subtopics, we used specific keywords including: adult attention deficit hyperactivity disorder, adult ADHD, adult ADHD diagnosis, adult ADHD medication, inattention disorder, hyperactivity disorder, oppositional defiant behavior, etc. Most of the posts collected were from early 2023. One researcher conducted a preliminary examination of the online posts to develop an initial set of criteria for labeling posts as either neutral or exaggerated. The research team then reviewed, discussed, and iterated on these criteria. The final set of criteria for selecting neutral ADHD-related posts required that posts meet **both** of the following conditions:

(1) The posts contain straightforward, specific descriptions of symptoms or personal experiences related to adult ADHD without promotional or suggestive language aimed at influencing reader perceptions.
(2) The content should be neutral in tone, avoiding emotionally charged language or exaggerated descriptions, aiming to inform rather than persuade.

The final set of criteria for selecting exaggerated ADHD-related posts required that posts meet **at least one** of the following conditions:

(1) The posts include emphasized ADHD symptoms in an amplified or sensationalized manner, often using alarming or urgent language.
(2) The posts include unverified data or statistics that depict ADHD as more widespread or severe than supported by typical clinical findings, reinforcing an exaggerated perception of the disorder's prevalence or impact.

For the control group, we selected social media posts from a diverse set of health communities. Hypotension, gastroesophageal reflux disease, aphantasia, insomnia, and chronic obstructive pulmonary disease were included due to their frequent presence and active discussions in online health forums. We then gathered and categorized these posts until all researchers agreed that information saturation had been reached, with no new content emerging. Each of the three categories ultimately included approximately 50 posts. We revised our categorization through a collaborative and iterative process.

*3.2.3 Content generation of social media posts.* We then combined representative expressions from each category and paraphrased them using language generation models to ensure consistency in rhetorical style across all posts. Original images from the posts were retained to maintain alignment between the visual content and the message conveyed in each post, enhancing the realism of the simulated platform. Virtual profiles and net names appeared in the posts were randomly generated with a diffusion model and manually checked by researchers to ensure that they were appropriate to the subjects.

As shown in Figure 1, the control group was exposed to five social media posts about health issues unrelated to adult ADHD. By including a variety of non-ADHD related health issues typically encountered by users on social media platforms, the control group served as a baseline without availability bias. The neutral group read posts that relatively accurate to reflect the nature of adult ADHD. The exaggeration group was exposed to content containing misleading descriptions of the prevalence and severity of the disorder, which unintentionally inflated perceptions of both.

---
[1]https://www.Reddit.com
[2]https://x.com
[3]https://www.Facebook.com

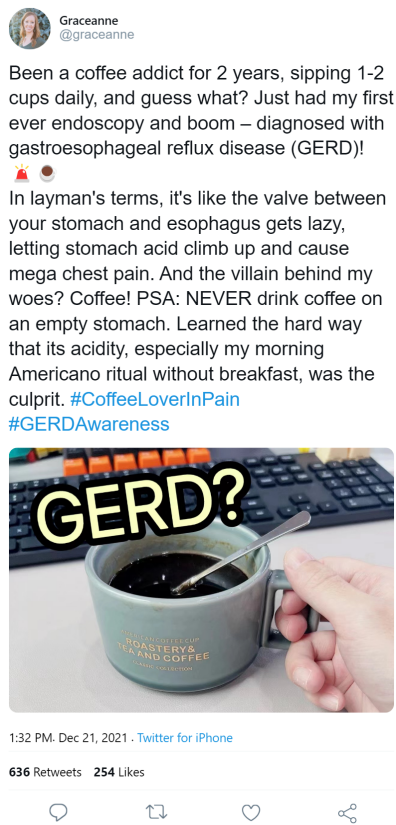
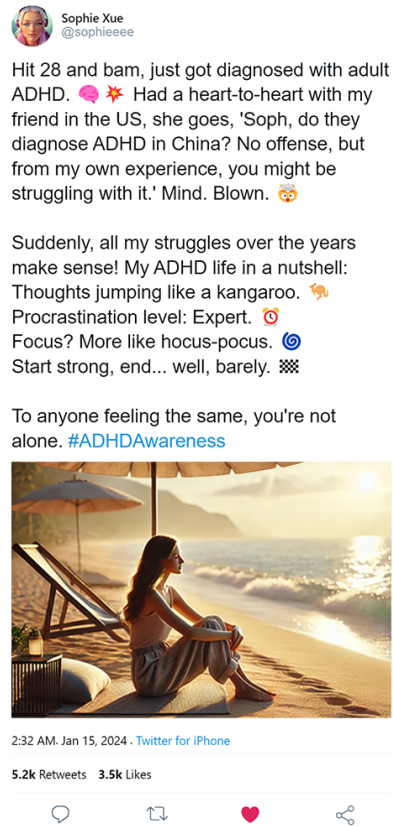
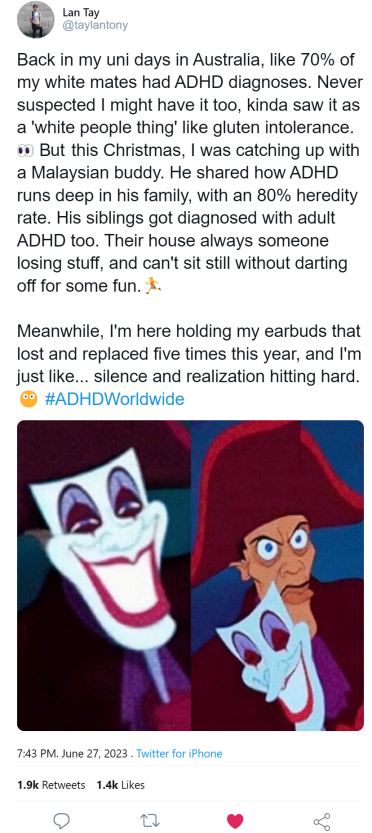

(a)                                                                                  (b)                                                                                  (c)

Figure 1: Simulated social media platform with three types of health-related posts: (a) controlled information, (b) neutral information, (c) exaggerated information.

### 3.3 Study Procedure

The study procedure includes four parts as shown in Figure 2, a pre-survey, the main task, the post-survey, and a debriefing session. After filling out the consent form and demographic information, participants were required to complete six questions for collecting their health literacy in the pre-task survey. They were then asked to rate prior performance through six questions on hyperactivity and attention deficit as their baseline ADHD level.

For the main task, participants were instructed to view 5 posts on the simulated social media platforms. This exposure phase was crucial for manipulating availability bias, so participants must correctly answer the questions about the details in each posts to proceed to the next task. After the exposure, participants then completed 18 questions to evaluate their adult ADHD levels and were provided with their evaluated outcomes with medical suggestions.

In the post-task survey, they reported the extents to which their self-assessment were influenced by social media. We collected their perceived information credibility through a questionnaire with eight items. Participants were also asked about the reason why their self-assessment were / were not affected by the social media contents. Before exiting the study, participants were debriefed on the purpose of the study and provided with knowledge about adult ADHD to correct misleading information mentioned in the posts.

### 3.4 Measurements

Quantitative outcomes, self-reported indicators, and qualitative comments were measured in this study to investigate the impact of social media exposure on participants' self-assessment.

*ADHD baseline:* The assessment of ADHD baseline were derived from a subset of question items extracted from SNAP-IV-26 scale [24]. Given that the purpose of this experiment was to explore the mechanisms by which social media exposure induces availability bias, blinding the participants was an important consideration. To prevent repeated measurements from revealing the experiment's purpose and to minimize potential memory effects [11], two different instruments, SNAP-IV-26 and ASRS, were employed in this short-term experiment. We selected two items for each of the three aspects that SNAP-IV-26 measures, which are inattention, impulsiveness, and oppositional defiant symptoms. Specifically, the item 3, 6, 17, 18, 22, and 24 were chose for less duplicative of the questions in the scale used in the main task. The internal consistency of the selected subscale was evaluated using McDonald's $\omega$ [28].

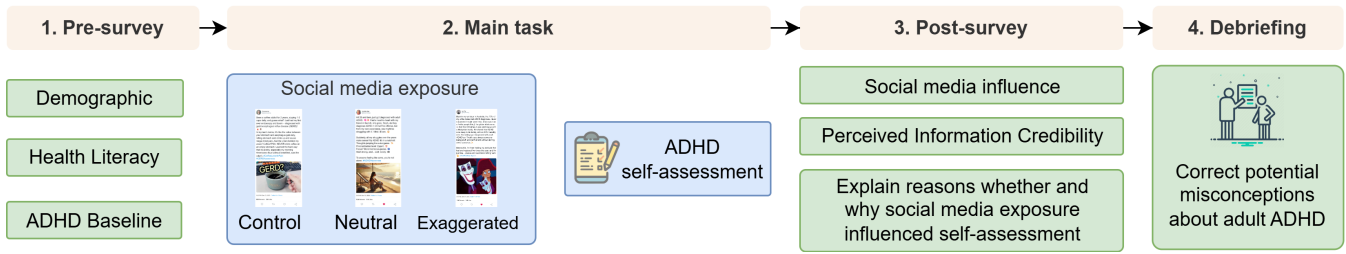

Figure 2: Flowchart of Study 1 procedure detailing the sequence of tasks and surveys. This diagram illustrates the comprehensive steps from the initial pre-survey, main tasks, post-survey and the debriefing session.

McDonald's $\omega$ yielded a point estimate of 0.79 (95% CI [0.73, 0.85]), indicating acceptable consistency. Same as the original scale's setting, answers were given on a 4-point Likert scale from "Not at all" (0) to "Very much" (3).

*Health Literacy:* Given that individuals' susceptibility to cognitive bias is related to health literacy [52], we chose this variable as a control variable. To measure individual's health literacy [4], the Newest Vital Sign (NVS) [88] was used. NVS scale consists of six questions that covers calculation and inference tasks designed to evaluate both numeracy and literacy skills in the context of health information. Each question is scored as either correct or incorrect, with the total score ranging from 0 to 6.

*ADHD assessment:* Since we aimed to investigate the diagnostic outcomes after social media exposure, we collected the evaluated ADHD score through the Adult ADHD Self-Report Scale (ASRS) [16]. This scale is used for the preliminary screening of ADHD symptoms in adults to determine whether further clinical evaluation is needed. 18 items in this scale provides a valid and comprehensive measure of ADHD symptoms covering both inattentive with 9 items and hyperactive-impulsive with another 9 items. We used a 5-point Likert scale ranging from "Never" (0) to "Always" (4) for each question. Both subtypes of ADHD symptomatology can be scored separately. Suppose that the baseline scores remain consistent across all groups, the difference in this score will indicate the extent to which different types of social media exposure can influence self-assessment outcomes.

*Social media influence:* We measured the extent to which participants' symptom self-assessments were influenced by social media and explored the connection between this influence and availability bias through follow-up open-ended questions. We measured this construct to assess each user's self-reported level of social media influence, reflecting the degree of availability bias. The evaluation was derived from Self-Reflection and Insight Scale (SRIS) [74]. We selected 7 questions from a subset of items in SRIS based on their relevance to our study's context and adapted them by adding "social media posts" to specify that they should consider the influence of social media exposure happens during their self-assessment. We also emphasized the self-assessment on their health conditions to clarify that we wanted participants to focus on the perception of the symptoms. The full scale can be found in Appendix.

*Perceived Information Credibility:* Since we aimed to investigate the influence of content with misleading information, we measured participants' perceived credibility towards the information provided in the social media content as manipulation checks. We adapted the questionnaire from an existing scale [23] by pointing out "the information in the posts" to specify they should consider their perception about the social media content. Participants were asked to rate on reliability, accuracy, trustworthiness, bias, and completeness on typical 5-point Likert scales.

*Explanation on the impact of social media exposure:* After the self-assessment, participants were asked, 'Do you think the posts you read influenced your self-assessment on your own symptoms of adult ADHD?' and 'Why do you think so?'. We use the answer to the first question to measure whether individuals perceived the impact of social media exposure, and the second question to obtain how the impact was caused by availability bias.

### 3.5 Participants

Potential participants were deemed eligible if they met all of the following inclusion criteria: (1) native English speaker located in the US; (2) above 18 years old; and (3) able to use computerized equipment to conduct surveys. The exclusion criteria were as follows: (1) previously diagnosed with ADHD (2) previously diagnosed with anxiety disorders, autism and other conditions that may lead to Attention Deficit Disorder (ADD). We conducted a power calculation for a three-group ANOVA study seeking a medium effect size with an alpha of 0.05. Given that N = 102 results in a statistical test power of 0.8, we recruited 104 participants (post-exclusion) on the Connect online crowdsourcing platform (CloudResearch [4]). Refer to Appendix on the breakdown of the demographic profile in each experimental condition. We ensured that the demographic profiles across the three conditions were similar so as to control for any fixed effects resulting from the differences in demographic factors. Participants who failed the pre-task attention check (2.8%) were excluded. Participants were randomly assigned to one of the three conditions. Each participant received $3.5 for an average task time of 22 minutes based on an hourly compensation rate of $10. This study was conducted with the approval of our local Institutional Review Board (IRB).

### 3.6 Data Analysis

To compare the outcomes of the four types of health information, we ran the analysis of covariance (ANCOVA) across groups. In each ANCOVA analysis, the independent variable was the type of social media exposure, categorized into four types, and the dependent variable was the measure specified in the hypothesis. Since research suggests that demographics and individual's health literacy influence people's susceptibility to availability bias [52], all analyses were controlled for participants' age, gender, educational level, and health literacy score. We additionally controlled ADHD baseline levels in the analysis of diagnostic outcomes as individuals' inherent symptom levels undoubtedly have a significant impact on the final diagnosis. After ANCOVA showed significance, we conducted the Tukey's post-hoc analysis to make pair-wise comparisons between conditions. Unless otherwise specified, all dependent variables in this study met the assumptions of ANCOVA, specifically homogeneity and normality tests. In this paper, we consider $p < 0.05$ to be statistically significant.

---
[4]https://connect.cloudresearch.com

For the qualitative data, we analyzed participants explanations about whether the social media exposure would affect their self-assessment process. Researchers first conducted thematic analysis based on whether they indicated there was influence in self-assessment or not and the reason behind. All responses were coded individually by two researchers, achieving inter-coder reliability with a Cohen's $\kappa$ of $0.84 > 0.80$, suggesting a high agreement. Any conflicts in coding were resolved through discussion.

## 4 STUDY 1: RESULT
### 4.1 Manipulation Checks

We compared perceived information credibility from post-survey responses to check the conditions using Kruskal-Wallis tests since the ratings were not normally distributed. Kruskal-Wallis tests indicated a significant difference was observed in perceived information accuracy ($\chi^2(2) = 25.9$, $p < 0.001^{***}$, $\varepsilon^2 = 0.27$) and trustworthiness ($\chi^2(2) = 19.3$, $p < 0.001^{***}$, $\varepsilon^2 = 0.17$) across groups. Post-Hoc analysis with Dwass-Steel-Critchlow-Fligner pairwise comparisons confirmed that the exaggerated group reported significant lower perceived information accuracy (Control: M = 3.59, SD = 0.75; Exaggeration: M = 2.85, SD = 0.77; $p = 0.008^{**}$), and lower perceived information trustworthiness (Control: M = 3.74, SD = 0.86; Exaggeration: M = 3.00, SD = 0.96; $p = 0.016^*$). The items which measured information trustworthiness and accuracy validated that *the exaggeration design of the posts was effective.*

### 4.2 Social Media Influence (H1.a, H1.b)

**Impact of neutral health information:** An analysis of covariance showed that the effect of social media exposure was significant ($F(2,98) = 4.21$, $p = 0.018^*$). Post-hoc analysis indicated that participants generally reported a greater influence of neutral content on their self-assessment of ADHD symptoms (Control: M = 19.71, SD = 7.61; Neutral: M = 25.79, SD = 9.38; $p = 0.22^*$, Cohen's D = -0.65). This result support **H1.a**: compared to social media users who were not exposed to shared experiences related to a disease, those exposed to neutral content were more influenced by social media exposure when self-assessing their symptoms.

**Impact of exaggerated health information:** From the Post-Hoc analysis, there was no significant difference between the exaggeration condition and the controlled condition (Exaggeration: M = 24.93, SD = 8.85, $p = 0.07$, Cohen's D = -0.55). There was no evidence supporting **H1.b**. This may suggest that exaggerated health information caused less impact on users' self-assessment of symptoms compared to neutral information.

### 4.3 Self-Assessed Diagnostic Scores (H2.a, H2.b)

*4.3.1 ADHD baseline level control.* To justify the influence of different conditions on the diagnostic outcomes, we first ensured the ADHD baseline level across groups were uniformly distributed. Specifically, Kruskal-Wallis test indicated there was no significant difference across conditions ($p = 0.72$) since the ADHD baseline level were not normally distributed. All participants reported a baseline level with mean values of 4.53 and standard deviation of 2.95. Distributions of participants' ADHD baseline levels are provided in Appendix.

This result indicates that individual differences in ADHD among participants across the three experimental groups were not significant.

**Neutral content induced overestimation of both inattention and hyperactivity symptoms.** From the ANCOVA analysis with the measured baseline levels as an additional covariance, a significant difference was observed on the ADHD inattention score ($F(2,98) = 9.95$, $p < 0.001^{***}$, $\eta^2 = 0.11$) and hyperactivity score ($F(2,98) = 3.90$, $p = 0.024^*$, $\eta^2 = 0.04$) across conditions. Post-Hoc analysis showed that the difference of inattention scores between the control group and neutral group is highly significant (Control: M = 13.37, SD = 7.08; Neutral: M = 18.30, SD = 8.14; $p < 0.001^{***}$, Cohen's D = -1.056). For the hyperactivity score, statistical significance was observed between the control group and neutral group (Control: M = 12.54, SD = 7.30; Neutral: M = 14.62, SD = 8.86; $p = 0.05^*$, Cohen's D = -0.58). The results are summarized in Fig. 3. These results support **H2.a**, showing that exposing to neutral health-related social media content causes an overestimation of self-evaluated symptoms.

**Exaggerated content did not led to overestimation of symptoms.** From the Post-Hoc analysis, no difference was observed between the controlled condition and the exaggeration condition either in inattention scores (Exaggeration: M = 15.78, SD = 6.88; $p = 0.42$, Cohen's D = -0.31) or hyperactivity scores (Exaggeration: M = 13.12, SD = 7.07; $p = 0.99$, Cohen's D = 0.04). The results indicate that participants who did not read posts related to adult ADHD reported symptoms at the same levels as those who read content exaggerating the prevalence and severity of adult ADHD symptoms. **H2.b** is not supported from this result.

### 4.4 Qualitative Results

To further answer **RQ1**, we gathered participants' explanations from the open questions on how social media posts influenced their self-assessment of adult ADHD across different conditions. We referred to any specific participant as P$x_y$, where $x$ is the participant number and $y$ represents different types of health information they were exposed to (*N* for neutral and *E* for exaggerated).

**Overall perceptions of the influence:** In the control condition, only a small minority of participants (3/34) believed that the social media content they read had an impact on their self-assessment of symptoms, which was expected as the posts were unrelated to adult ADHD symptoms. In the neutral condition, the majority of participants (24/35) reported being influenced by the social media posts. In the exaggeration condition, a relatively small portion of participants (9/35) indicated that the posts affected their self-assessment. This result aligns with the previous findings on social

Table 1: The mean and standard deviation for each group regarding perceived information accuracy, perceived information trustworthiness, social media influence, ADHD inattention score, and ADHD hyperactivity score, along with the p-values for intergroup differences.

| Variables | Group | | | p value | | |
|---|---|---|---|---|---|---|
| | Control (C) | Neutral (N) | Exaggerated (E) | C v.s. N | C v.s. E | N v.s. E |
| **Perceived information accuracy** | 3.59 (0.75) | 3.96 (0.96) | 2.85 (0.77) | 0.10 | 0.004** | <0.001*** |
| **Perceived information trustworthiness** | 3.74 (0.86) | 3.89 (0.92) | 3.00 (0.96) | 0.74 | 0.009** | 0.002** |
| **Social media influence** | 19.71 (7.61) | 25.79 (9.38) | 24.93 (8.85) | 0.022* | 0.07 | 0.91 |
| **ADHD Inattention Assessment** | 13.37 (7.08) | 18.30 (8.13) | 15.78 (6.88) | 0.012* | 0.53 | 0.16 |
| **ADHD Hyperactivity Assessment** | 12.54 (7.30) | 14.62 (8.86) | 13.12 (7.07) | 0.47 | 1.00 | 0.47 |

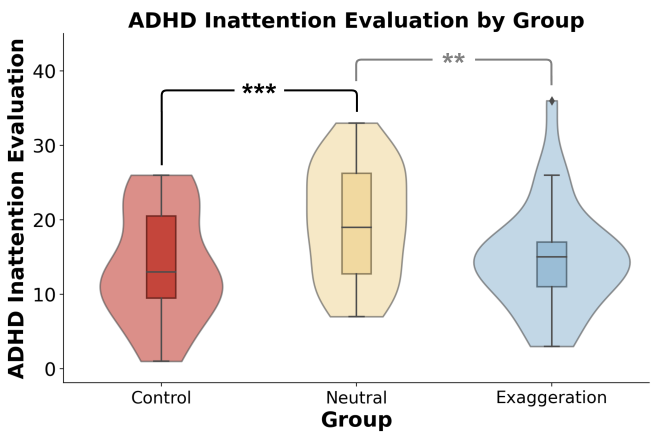

(a) Evaluated scores of adult ADHD-inattention sub-type.

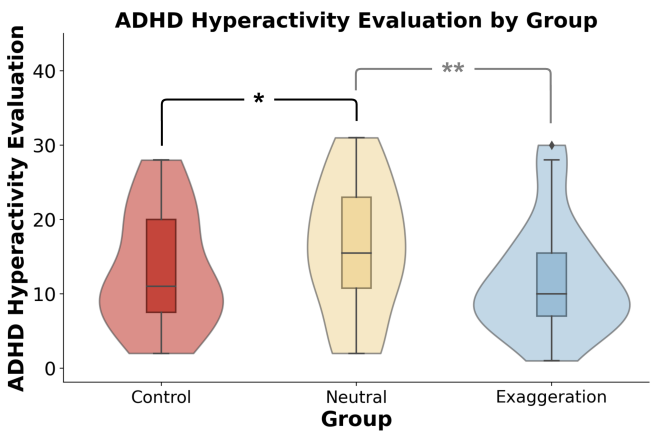

(b) Evaluated scores of adult ADHD-hyperactivity sub-type.

Figure 3: Participants' ADHD self-assessment scores from post-survey across groups. (*p < 0.05, **p < 0.01, ***p < 0.001)

media influence, further confirming that neutral information had the greatest impact.

We further conducted a thematic analysis of the reasons participants were influenced or not, aiming to uncover the connection between this influence and availability bias. Researchers coded the responses from all participants who were not influenced, categorizing them into three main reasons: (1) I based my judgment entirely on my own facts, (2) I didn't relate my experiences to that shared in the posts, and (3) I don't think those posts are credible. For the reasons why self-assessments were influenced by social media content, we also identified three explanations: (1) The experiences in the posts are very close/related to mine, (2) The posts emphasized similar experiences in my mind, and (3) I compared my symptom levels with those described in the posts. Although an individual may explain multiple reasons, since nearly all participants provided only one primary reason, we instructed the researchers to assign each response to just one main category.

**Neutral health information triggers availability bias by evoking resonance.** We found that the primary reasons of being influenced were the relevance of the post content to own experiences (8/24), and the emphasis on symptoms (6/24) in the neutral group. Participants who had previously experienced issues like inattention, losing things, or procrastination saw themselves reflected in these posts. In these cases, participants felt they shared some common experiences with adult ADHD patients, and this sense of similarity led them to believe that those symptoms frequently applied to themselves as well. As one stated:

> "I found that I could relate with some of the details included in the posts, I also struggle with things like focusing on work activities and articulating a point clearly to others. I recognized some of those symptoms in the self-assessment and I made sure to mark those higher..." (P67$_N$)

Another commonly reported reason from participants was that repeated exposure to information about ADHD symptoms emphasized the perceived frequency of its occurrence in their minds. They reported that, having just been exposed to similar information, experiences related to negative health conditions were more easily recalled. One participants called,

*"[t]he content presented was quite fresh on my mind right before answering the questions so I reflected more on the time I lost attention on my work and made some mistakes..."* (P85$_N$)

**Exaggeration did not induce more availability bias compared to the neutral social media exposure.** In contrast to neutral content, exposure to exaggerated social media posts appeared to have less impact on participants' self-assessment. A significant number of participants (11/26) indicated that the exaggerated nature of the posts made it difficult to relate the content to their own lives. As one participant noted:

*"The tweets had nothing to do with the reality of what I experience. I read the tweets and answered the questions. When assessing my own experience, I didn't feel the need to reflect on the Twitter posts, therefore they did not influence my answers."* (P44$_E$)

Moreover, some participants noted that the exaggerated portrayal of symptoms in the posts led them to question the validity of the content. For these individuals, the posts appeared sensationalized, and they were skeptical of their applicability to real-life situations. This skepticism further reinforced their decision to disregard the posts when assessing their own behaviors. As one participant commented:

*"I feel like most people don't experience anything this serious in reality, these posts seem a bit sensationalized."* (P46$_E$)

Overall, the majority of participants either maintained a strong focus on their own facts, dismissed the content as unrelated to their experiences, or distrusted the posts. These findings suggest that the impact of exaggerated social media posts on self-assessment appears to be minimal compared to the neutral content, as participants were more likely to resist its influence due to skepticism and a lack of personal resonance.

## 4.5 Summary of the Findings

We observed a trend from both the qualitative and quantitative results where social media exposure to neutral health information significantly influenced participants' self-assessment of ADHD symptoms. The neutral health information evoked availability bias through resonance with individuals' personal experiences and emphasizing the occurrence of symptoms. The impact of this bias was evident in both self-reported social media influence and their diagnostic outcomes.

In contrast, the exaggerated content seemed to help these participants differentiate between their own life experiences and the extreme symptoms depicted in the posts. This discernment protected them from the potential influence of availability bias, as they recognized that the posts did not accurately reflect their personal reality. This differs from our expected results, but as indicated by the qualitative findings, susceptibility to availability bias is more closely related to the relevance of social media content to the users' own experiences.

Overall, the findings suggest that resonant social media content triggered availability bias, leading users to overestimate their symptom levels during online self-diagnosis.

## 5 STUDY 2: COMPARING THE EFFECTS OF CHATBOT-BASED SYMPTOM CHECKERS

Study 1 demonstrates that social media exposure induces availability bias by causing users to disregard their own evidence and misattribute others' experiences in their minds. Prior literature has shown that chatbots have the potential to promote evidence-based self-reflection [19, 77], which help individuals resist cognitive biases through reflective thinking patterns [15, 48, 49]. Hence, we designed cognitive intervention strategies targeting the primary causes of availability bias found in Study 1 and implemented them in CSCs to promote evidence-based self-reflection.

### 5.1 Conditions

To answer **RQ2**, we kept the neutral condition same as Study 1, which induced the most significant availability bias. All four conditions were provided with the same information (i.e., images and texts) as the neutral social media posts presented in Study 1. Through a mixed-method study (N=100), we compared the effectiveness of three chatbot designs in mitigating availability bias. Fig. 4 shows examples of how different CSCs interact with the user. All prompt settings for GPT-4 are listed in the Appendix.

*5.1.1 Control: Static Questionnaire.* Self-assessment in the control group is similar to the experience with conventional OSCs [72]. The users were asked a series of questions about their symptoms shown on a static web page. After they complete filling out the questionnaire, the system will provide an diagnostic result with medical suggestions.

*5.1.2 CSC.* Apart from cognitive intervention strategies, we also aimed to investigate the role of the chatbot itself in reducing availability bias since natural dialogues can elicit self-reflection on new evidence or different perspectives [36, 90]. In this condition, we replaced static web-based survey with a chatbot-based symptom checker without cognitive interventions. We developed a CSC with doctor-like probing mode and emotional support based on the design proposed by You et al. [92]. To be specific, we applied friendly addresses, greetings, a small number of caring words, encouragement, and potential diagnosis suggestions during each conversational stage.

*5.1.3 CSC with Evidence Reflection.* This design was on the basis of the CSC condition's settings and was inspired by works of Schwarz et al. [71] and Tetlock and Kim [78], who have proposed when people need to explain and defend their judgments and decisions, they would consider facts and evidence more carefully. Given that resonant health information leads users to misapply others' experiences to them, the CSC with Evidence Reflection strategy was designed to question closely about the circumstances in which their symptoms occurred and the specific details of those symptoms, until the user provide sufficient context and explanation. It was prompted to first judge whether the user has ever experienced certain health condition from the conversation, if yes, it would further analyze whether the response mentioned a specific evidence.

*5.1.4 CSC with Counterfactual Thinking.* This design was also built on the CSC condition. The results of Study 1 indicated that availability bias emphasizes events involving negative conditions in users' minds, leading them to overlook experiences where the disorder did not occur, which echoed with Davidai and Gilovich's [18] findings. The counterfactual design was inspired by what Kahneman [34] and Johnson [32] has observed, counterfactuals are effective in encouraging thinking from multiple perspectives and thus disrupt heuristic thinking. In this condition, users are required to consider the frequency and circumstances in which a symptom does not occur when they indicate that they have ever experienced a symptom, viewing it from the opposite perspective.

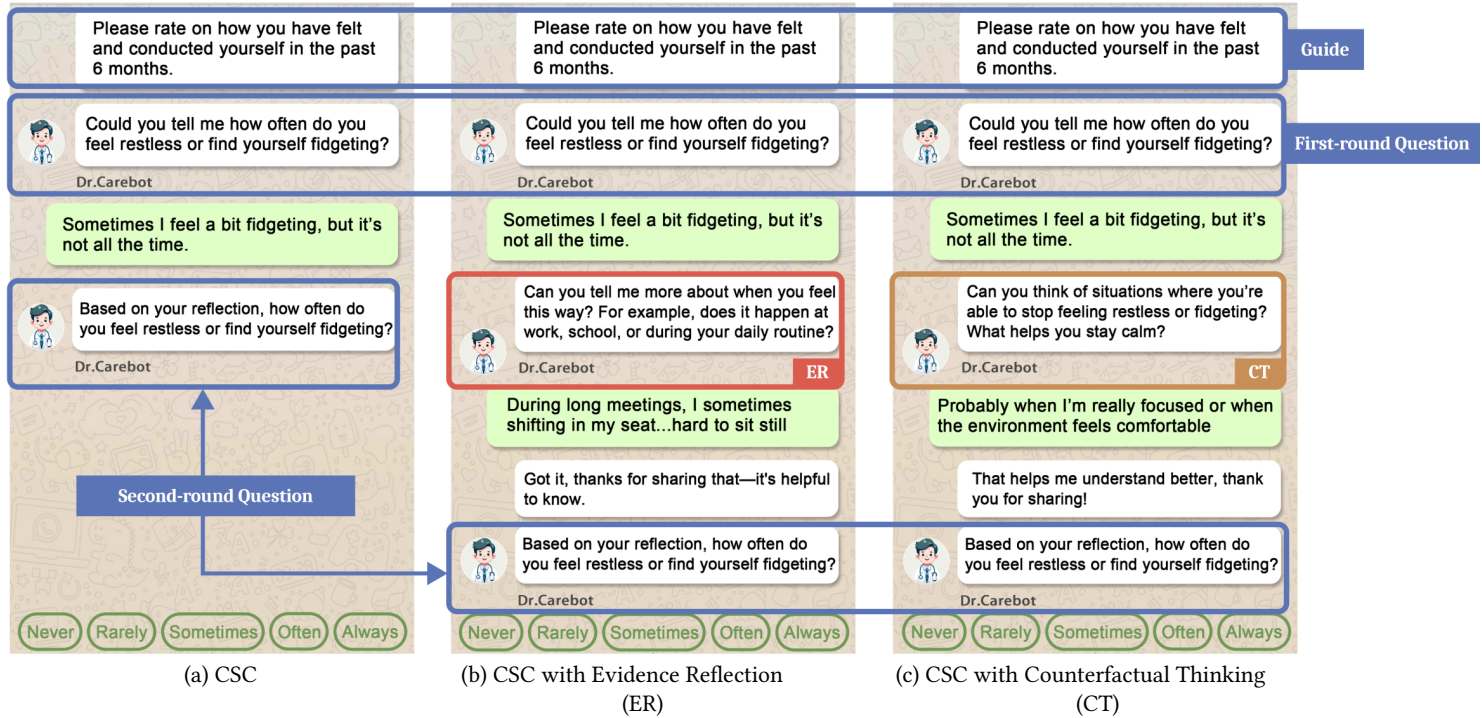

Figure 4: Examples of symptom checking questions asked by the designed CSCs. The blue sections represent controlled explanations in responses. The red and yellow sections are uniquely crafted cognitive interventions, aimed at encouraging deeper reflection.

## 5.2 Study Procedure

The procedure of this study is shown in Fig. 5. After obtaining participant's consent and demographic, we first collected their baseline level of adult ADHD with ASRS-v1.1 scale. During the social media exposure, participants went through five posts from the neutral group in Fig. 1, which showed the most significant effect on inducing availability bias in Study 1. Participants also completed detailed comprehension questions about these posts to make sure they had read the contents carefully.

In the main task, participants were instructed to do a self-assessment on adult ADHD with a conventional OSC or the designed CSC. The four conditions in this study were between-subjects, i.e., participants were randomly assigned to one of the experimental. The assessed scores from the main task were as within-subjects variables to be compared with participants' baseline levels and were reported to demonstrate whether their diagnostic outcomes were affected.

After the participants were done with the main task, they filled out post-survey questionnaires to self-report the extents to which they were influenced by social media and mental effort they put into self-reflection. Additionally, participants were asked to provide qualitative comments on why the social media content changed/did not change their perception and how interacting with the chatbot affect the way of self-reflection on their health conditions. Finally, a debriefing session is provided to correct possible misunderstanding from the social media posts and emphasized the medical suggestions from this study was for informational purposes only and does not represent any trustworthy medical guidance.

## 5.3 Measurement

This study involves three quantitative measurement: ADHD self-assessment outcome, social media influence, and mental effort during self-reflection. Basic demographic information, e.g., age, gender, education, and general experiences of using chatbots were collected at the beginning.

*ADHD assessment:* Since we aimed to compare the change of diagnostic outcomes, we measured participants' baseline adult ADHD level and post adult ADHD level in both pre-survey and the main task. The questionnaire was adapted from Adult ADHD Self-Report Scale (ASRS-v1.1) [16]. We kept the original questions of ASRS-v1.1 for measuring participants' baseline data in the pre-survey and changed some of the rhetoric in the second measurement to fit the conversational nature and avoid repetition.

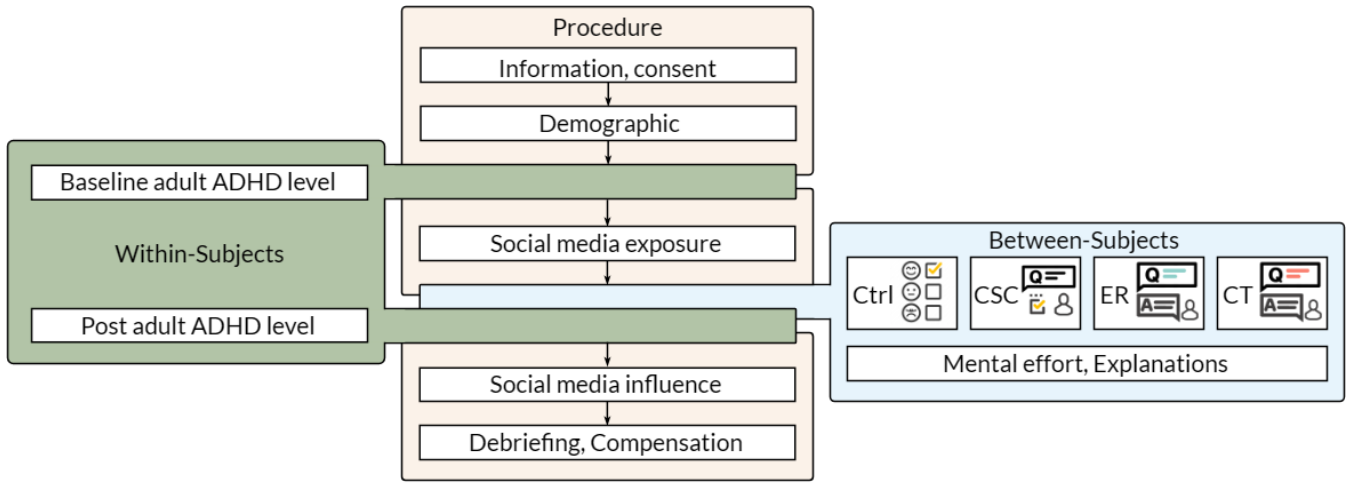

Figure 5: Overview of the experimental procedure.

*Social media influence:* The evaluation of social media influence followed the design in Study 1. 7 questions from SRIS was selected to measure the influence of social media on self-assessment as shown in Appendix.

*Mental effort:* Depth of self-reflection undertaken by the participant was measured by NASA-Task Load Index (NASA-TLX) [27]. We adapted two questions to quantify mental effort during self-reflection: "*How much mental and perceptual activity was required (e.g., thinking, deciding, calculating, remembering, looking, searching, etc.)?*" and "*How hard did you have to work mentally to accomplish the self-assessment?*" One item for measuring temporal demand: "*How much time pressure did you feel due to the rate or pace at which the tasks occurred?*", and one item for measuring perceived performance: "*How successful do you think you were in accomplishing the goals of self-assessment?*".

*Explanation on the impact of social media exposure and the chatbot interaction:* A qualitative follow-up were conducted after the study. We asked all participants to indicate whether the social media content they read would affect the reflection on their own health conditions and explain the reason. Participants in the treatment groups were asked to answer whether interacting with the chatbot changed the way they self-reflected compared to when they first did the static questionnaire. For each question we enforce a response of more than 50 characters.

### 5.4 Participants

We conducted a power calculation for a four-group ANOVA study seeking a medium effect size, at 0.80 observed power with an alpha of 0.05, given N = 24 per experimental condition, hence we recruited 100 participants in total. The inclusion and exclusion criteria were the same as in Study 1. Of the 132 participants who started the study, 100 completed the study and passed our attention check. Our analysis is based on those 100 valid responses with the complete demographic information available in Appendix. As for their familiarity with chatbot technology, 77 out of 100 participants reported that they have used ChatGPT or other chatbot-based products before, and all of them feel comfortable using chatbots in general. The study lasted approximately 25 minutes, with participants earning an average wage of about $10 per hour. The study procedures were approved by our local Institutional Review Board (IRB).

### 5.5 Data analysis

To compare the effects of different cognitive strategies with CSCs between subjects, an One-Way ANOVA was conducted for dependent variables which conforms to the normal distribution and equal variance assumptions. The exception was temporal demand, which was tested with Kruskal-Wallis. If the One-Way ANOVA showed significance, Tukey's post-hoc analysis was used to compare difference between conditions. We also compared the change in the self-evaluated outcomes of adult ADHD before and after the intervention. After confirming that the difference between the scores from pre- and post-survey for each condition satisfies a normal distribution, we ran the Paired Samples T-Test to examine their differences.

The qualitative data obtained from the open-ended questions were coded separately by two researchers followed deductive coding patterns achieving inter-coder reliability with a Cohen's $\kappa$ of $0.88 > 0.80$ (high agreement). Any conflicts in coding were resolved through discussion.

## 6 STUDY 2: RESULT

As the prompts used in the chatbot development, the AI agent adhered to our instructions. In the CSC with Evidence Reflection, the chatbot effectively bypassed requests for further explanation when the user's response provided sufficient context. In the CSC with Counterfactual Thinking design, the chatbot consistently identified the user's inclination regarding whether they had experienced a particular symptom and accurately posed counterfactual questions.

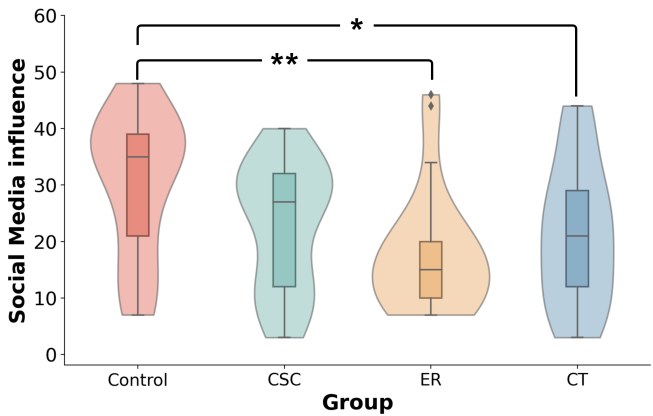

Figure 6: Comparison of Social Media influence scores across different conditions, including CSC with Evidence Reflection (ER), CSC with Counterfactual Thinking (CT), and the control group. Statistical significance is denoted by p-value annotations, where *: p < 0.05, **: p < 0.01.

From participants' feedback, the vast majority of them perceived the conversation to be natural and intuitive.

### 6.1 Social Media Influence

A one-way ANOVA was conducted to examine the effect of different media on influence scores. Result indicated a statistically significant difference between groups in the influence of social media exposure on self-assessment ($F(3,96) = 5.10$, $p = 0.003$**, $\eta^2 = 0.14$). A Tukey HSD post hoc test was conducted to explore the differences between conditions. The results indicated that participants who used the CSC with Evidence Reflection strategy (M = 17.80, SD = 10.62) reported a significantly lower score than those who used traditional questionnaires (M = 30.40, SD = 12.48), $p = 0.001$**, Cohen's D = 1.07. Similarly, the CSC with Counterfactual Thinking condition (M = 21.32, SD = 12.60) scored significantly lower than the control condition, $p = 0.037$*, Cohen's D = 0.77. Participants who used the CSC (M = 22.88, SD = 11.19) did not report a statistically lower influence induced by the social media compared to the control group, $p = 0.11$, Cohen's D = 0.64. The scores for different groups are visualized in Figure. 6. These findings suggest that the use of CSCs with cognitive interventions significantly reduced the influence of social media compared to traditional questionnaires.

### 6.2 Self-Assessed Diagnostic Scores

Diagnostic scores across all groups are shown in Fig. 7 with the significance from paired samples t-test labeled. We compared inattention and hyperactivity scores before and after social media exposure.

There was a significant increase in inattention scores from baseline (M = 13.40, SD = 5.28) to post-evaluated score (M = 15.60, SD = 6.32), $t(24) = -3.29$, $p = 0.003$**, with a medium effect size (Cohen's d = -0.66). This result again proved that social media exposure would cause an overestimation on inattention symptom levels.

For the CSC, the result indicated that there is still a significant increase in inattention scores following the social media exposure, with a shift from baseline (M = 12.12, SD = 6.43) to post-evaluated score (M = 13.08, SD = 5.90), $t(24) = -2.72$, $p = 0.012$*. The effect size was moderate (Cohen's d = -0.54). This suggests that this treatment group still overestimated their inattention symptoms. However, the smaller effect size under this condition the treatment was somewhat effective in mitigating the overestimation in inattention symptoms.

For the CSC with Evidence Reflection, there was no significant change in inattention scores from baseline (M = 14.56, SD = 6.91) to post-evaluated score (M = 15.32, SD = 7.40), $t(24) = -1.05$, $p = 0.31$, with a small effect size (Cohen's d = -0.21). For those who used CSC with counterfactual strategy, the change in inattention scores from baseline (M = 15.12, SD = 7.78) to post-evaluated score (M = 15.44, SD = 7.46) was not statistically significant, $t(24) = -0.37$, $p = 0.72$, with a very small effect size (Cohen's d = -0.07). Across all four groups, there were no statistically significant differences in hyperactivity levels before and after social media exposure. This indicates that both treatments with cognitive strategy was effective in addressing the overestimation of inattention scores due to availability bias.

### 6.3 Mental Effort

Mental effort during self-assessment under the four conditions is shown in Fig. 8. A one-way ANOVA using Welch's correction was conducted to examine the effect of different groups on mental effort they engaged on self-assessment. The analysis revealed a statistically significant difference between groups, $F(3, 53.3) = 4.37$, $p = 0.008$**. Post-hoc comparisons using the Tukey HSD test indicated that participants using CSC with the Counterfactual strategy reported significantly higher total mental effort compared to the control group (control: M = 6.80, SD = 3.03; CSC with Counterfactual Thinking: M = 9.56, SD = 2.95; $p = 0.007$**). The CSC with Evidence Reflection condition also had significantly higher total mental effort than the control group (CSC with Evidence Reflection: M = 9.20, SD = 2.72; $p = 0.024$*). No significant differences were found between the other group comparisons (all $p > 0.05$).

This result suggests that CSCs with cognitive intervention strategies encouraged users to invest more mental effort in self-reflection, thereby mitigating the negative impact of availability bias.

### 6.4 Qualitative Results

**Effects of cognitive interventions:** We gathered qualitative insights on the effectiveness of CSC designs, compared to static questionnaires, in facilitating self-reflection. Two researchers first coded 75 responses from the three groups that used the CSC, categorizing them into two groups: those who noted differences in using the CSC for adult ADHD self-assessment compared to the questionnaire, and those who did not. In the CSC condition, 10/25 participants reported that the chatbot prompted a different thinking pattern. In the CSC with Evidence Reflection and Counterfactual Thinking groups, this proportion was 15/25 and 13/25, respectively. Next, we focused on participants who reported differences and conducted further coding to analyze the reasons they provided. Results from thematic analysis revealed three main categories of differences: (1) increased cognitive effort and time, (2) greater recall of specific examples, and (3) a sense of being listened by a real person. Responses were allowed to fall into multiple categories, as many participants provided detailed explanations that addressed more than one way in which the CSC differed from a static questionnaire.

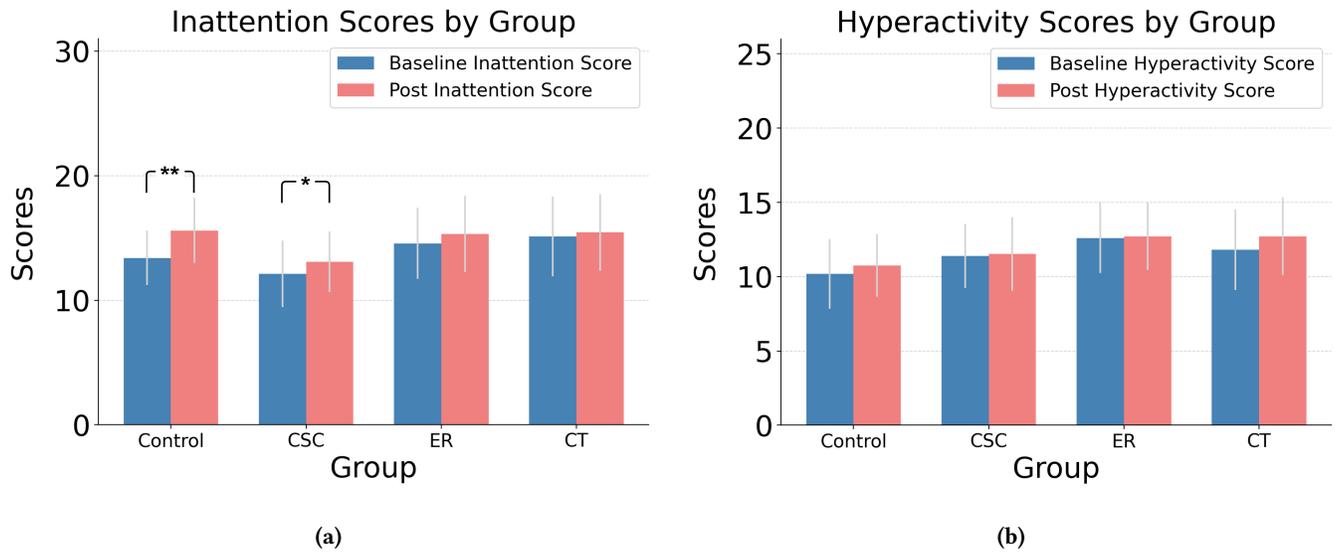

Figure 7: Baseline and post-intervention inattention (a) and hyperactivity (b) scores across groups. Significant increases in inattention scores were found in the Control (p < 0.01) and CSC (p < 0.05) groups. ER and CT refers to CSC with Evidence Reflection and CSC with Counterfactual Thinking. Error bars show .95 confidence intervals. Statistical significance is denoted by p-value annotations, where *: p < 0.05, **: p < 0.01.

Table 2: The mean and standard deviation for each group regarding social media influence and mental effort.

| Variables | Group | | | | p value | | |
|---|---|---|---|---|---|---|---|
| | Control (C) | CSC | ER | CT | C v.s. CSC | C v.s. ER | C v.s. CT |
| Social media influence | 30.40 (12.48) | 22.88 (11.19) | 17.80 (10.62) | 21.32 (12.60) | 0.11 | 0.001** | 0.037* |
| Mental effort | 6.80 (3.03) | 8.00 (3.03) | 9.20 (2.72) | 9.56 (2.95) | 0.47 | 0.024* | 0.007** |

In the group that applied the CSC with Evidence Reflection cognitive intervention strategy, 7/15 participants reported spending more time and effort on reflection. Participants generally felt that responding to the chatbot's questions in words required more careful thought compared to simply "checking a box" or "picking a bubble" in a questionnaire. For instance:

> "The chatbot gave me more time to actually think about it and reflect on the actual answers instead of just picking a bubble. I think there were some changes in my answers after more thought." (P65$_{ER}$)

Another common experience reported in the CSC with Evidence Reflection condition was that users needed to describe specific examples related to particular symptoms to the chatbot. This process also compelled them to actively recall details based on their own experiences. As one participant commented:

> "The chatbot ... makes me recall it as I'm thinking of specific examples to illustrate the point, instead of static checking boxes you're more thinking of the subject and the frequency, and typically have an inner scale to select what you feel." (P51$_{ER}$)

In the CSC with Counterfactual Thinking condition, fewer participants (2/13) mentioned they reflected more on evidence-based examples, as this cognitive intervention did not require such examples. A significant number of participants (5/13) still reported that they engaged in deeper self-reflection compared to completing a static questionnaire. Several participants noted that the Counterfactual design prompted them to consciously differentiate between manageable behaviors like inattention and actual disorders happen on patients. As one participant noted:

> "It made me actually think about what is like having ADHD and what surmountable difficulties in life are not symptoms of ADHD... I think I reflected more deeply on what each question meant." (P76$_{CT}$)

Overall, the cognitive intervention strategies in CSCs promoted self-reflection, helping users to focus more on their specific experiences and think comprehensively. This enabled them to engage in reflective thinking, thereby mitigating the negative effects of availability bias.

**Effects of conversation-based self-diagnose:** Apart from the effects of cognitive interventions, some participants using the CSC without these strategies reported that the way they thought during self-assessment in conversation with the chatbot differed from when completing a questionnaire. In fact, across all three treatment groups, some participants (11/39) reported feeling as though they were conversing with a real person while using the CSC. This gave participants a sense of being actively listened to. This feeling encouraged them to open up more and prompted them to engage in deeper thinking in order to provide more detailed responses. As two participant explained:

> "I have chance to describe my experience instead of simply ticking the checkbox. It made me more willing to pour out what I suffered and now I feel I know my situation better." (P36$_{CSC}$)

> "I really enjoyed how it seemed like a conversation rather than boring questions that ask how I view myself. It made me want to open up more about what I experience because I felt like someone was actively listening to me." (P67$_{ER}$)

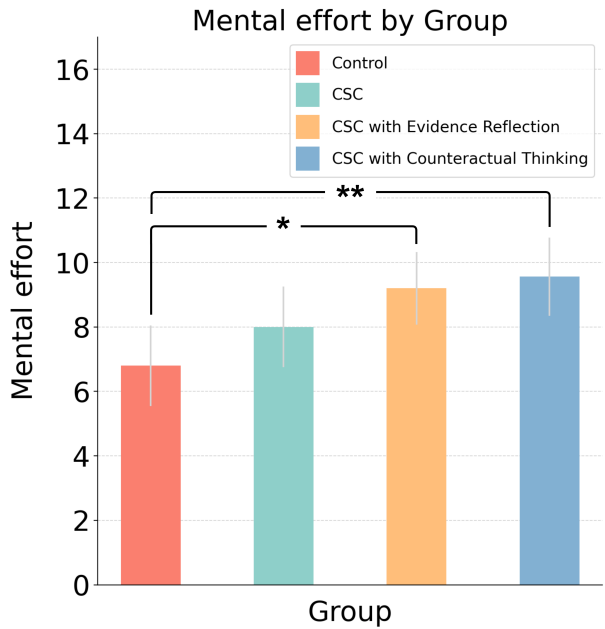

Figure 8: Mean evaluation scores on mental effort across the four conditions for study 2. Error bars show .95 confidence intervals. We report the results of the one-way ANOVA test and pairwise comparisons, where * : p < 0.05, ** : p < 0.01.

These quantitative results suggest that the increased engagement when describing personal experiences to the chatbot may encourage active thinking, potentially facilitates self-reflection.

**Negative comments:** In the collected qualitative feedback, a small number of participants (6/75) in each group who used the CSC provided negative comments. They expressed that the conversational format was, in some aspects, less usable than static questionnaires. One of the dominant complaints was the perceived inefficiency of the chatbot. As one participant noted,

> "I was mostly annoyed that the chatbot asked the same questions as the questionnaire but it took 10x times." (P99$_{CT}$)

Another recurring theme is the perceived redundancy and lack of meaningful change or adaptation in the chatbot's interaction. Some participants noted that the chatbot merely replicated a programmed function without adding value, which was described as *"more stressful than helpful"*. These negative feedback suggests that the chatbot's dynamic interaction, which is intended to enhance user engagement, may in fact detract from the experience when the technology fails to perform optimally.

## 7 DISCUSSION

### 7.1 Resonant Social Media Content Induced Availability Bias

Regarding **RQ1** on how health-related information on social media triggers availability bias in symptom self-assessment, we found that neutral content leads users to overestimate their symptom levels since the relevance caused them to disregard their own evidence. This finding **supports the hypothesis (H1.a, H2.a)**, and aligns with existing literature in the healthcare domain [9, 30, 43, 50, 94], which indicated that frequent exposure under health information could trigger availability bias in healthcare context. Our findings extend the prior literature by showing that resonated health-related information induces availability bias by making certain symptoms more readily recalled and leading them to believe that symptoms experienced by others also applied to them. These results should be taken into account when considering designing online tools for user's self-diagnosis.

However, contrary to the hypothesis regarding the exaggerated health information, the results indicate that it did not trigger more availability bias compared to neutral information. This finding **refutes the hypothesis (H1.b, H2.b)**. Based on participants' feedback, this can be explained by the fact that the exaggerated symptoms and prevalence in the health content did not closely align with their actual experiences. Also, the skepticism towards the credibility of misleading information may prevent them from being influenced by such content when assessing their symptom levels. This finding appears to extend previous literature [57] by suggesting that, in the context of self-diagnosis, it is likely the relevance of health information that may influence susceptibility to availability bias.

Interestingly, the results of Study 1 indicated that social media posts with exaggerated health information triggered less availability bias than neutral content. However, this does not imply that exaggerated information is beneficial, as it can cause stigma against patients [55] and provoke unnecessary panic [59]. Additionally, in this experiment, neutral and exaggerated content were isolated to understand their unique influences on symptom perception. In real-world scenarios, however, individuals are likely to be exposed to mixed content of both neutral and exaggerated health information on social media. Previous research highlights that exaggerated health information tends to spread faster and exert a greater influence on public perceptions compared to factual content [85]. This suggests that users exposed to both neutral and exaggerated health information may respond more similarly to the exaggerated condition in our experiment. Future research is needed to explore how these dynamics shape symptom perception in real-world contexts.

## 7.2 Cognitive Interventions in the CSC Reduced the Impact of Availability Bias

Study 1 showed that exposure to social media triggers availability bias in online self-diagnosis by causing them to disregard their own evidence during self-assessment. With regard to **RQ2** about the extent to which CSCs mitigate the impact of availability bias, Study 2 demonstrates the effectiveness of CSCs with cognitive intervention strategies. Our participants recognized the value of using CSCs incorporated cognitive intervention strategies to foster evidence-based self-reflection. This may prevent users from availability bias, leading to a reduced tendency to overestimate their symptom levels. This result resonates with existing research [10, 14, 20, 79] that found cognitive intervention to be an efficient strategy for bias mitigation.

The CSC with Evidence Reflection mitigates availability bias from social media exposure by prompting users to provide concrete symptom accounts, encouraging deeper analytical thinking and cognitive processing. These results are roughly in line with previous research [86] that the shift from heuristic recall to evidence-based reasoning disrupts the recall of recent or vivid events, thereby reducing availability bias. The effectiveness of the Counterfactual design lies in its ability to encourage users to think from multiple perspectives, prompting them to consider both the occurrence and non-occurrence of symptoms. This counterfactual reasoning broadens the range of scenarios considered, disrupting heuristic shortcuts and reducing availability bias, which is consistent with prior [32, 34]. The implications of these findings are significant for the design and deployment of CSCs in digital health contexts.

Despite these advantages, as described in Section 6.4, cognitive interventions in CSCs promoted self-reflection at the cost of increased redundancy. As a result, these designs may have lower adoption rates compared to the quicker, more straightforward traditional OSCs. We believe this limitation can be partially overcome by developing adaptive strategies [7]. In these strategies, the most effective cognitive interventions would be deployed selectively, targeting individuals most susceptible to availability bias and focusing on disorders with the most extensive recent social media exposure. Developing such adaptive strategies may require future study to work on the identification of susceptible populations to availability bias and monitor of health information on social media.

## 7.3 Conversational Nature of CSCs Has the Potential to Encourage Active Self-Reflection

To further address **RQ2** regarding the role of chatbots in promoting self-reflection, we focused on the qualitative feedback particularly about the chatbot itself. Our findings indicate that the conversational nature of the CSC prompted participants to engage in more active, thoughtful responses to the agent's questions. This result supports recent findings in the HCI community regarding the effectiveness of using Questioning-and-Answering frameworks to promote analytical thinking and logical reasoning [17, 58, 77]. As described in Section.6.4, the possible cause is a two-way process: users feel an obligation to provide the chatbot with more evidence to help it understand their health status, and they also feel actively listened to when confiding in the chatbot. The observed obligation is in line with previous research [60], suggesting that people mindlessly apply interpersonal social norms and expectations to computer agents since they treat them as social actors. Moreover, the feeling of being actively listened to, seem to echo Lee et al.'s [40] findings that users greatly appreciated the chatbot's encouragement to reflect on their health status. This observation suggests that the social dynamics between users and chatbots play a crucial role in enhancing self-reflection.

## 7.4 Design Implication

Our findings indicate that when users are exposed to health-related content on social media that resonates with them, they are likely to be influenced by availability bias and tend to overestimate their symptom severity. This is especially concerning in the digital age where users' health information sharing and seeking are becoming increasingly common [96], and the impact of social media in shaping public health behaviors is expected to continue growing. In light of these findings, it is crucial for future research to explore the development of both cognitive interventions within digital health tools and the mechanisms by which social media recommendation systems disseminate health-related content.

The effectiveness of our cognitive interventions in CSCs provides potential strategies for future online diagnostic tools. This approach highlights the role of cognitive interventions in supporting evidence-based self-reflection by guiding reflective reasoning, thus mitigate the impact of availability bias during the assessment process. Our experiment demonstrates an opportunity for new OSC design that guide the user to assess their health conditions through Questioning-and-Answering framework rather than directly reporting symptom levels [91]. We envision numerous future applications for integrating cognitive intervention strategies to chatbots, such as mitigating echo chamber effects of social media [77], or in decision-support scenarios for doctors [14], where cognitive biases can have critical consequences.

A key implication for the design of social media platforms is that algorithms that persistently recommend certain disease-related content may amplify users' concerns about those conditions, as our study 1 shows. Unlike personalized recommendations for other types of content, health information carries unique risks and consequences that require careful consideration. Prior studies has shown the importance of balancing user satisfaction with the risk of misleading health information dissemination when designing recommendation algorithms [64]. We encourage social media platforms to balance the delivery of health-related content with reminders that guide users to reflect on the evidence about their health conditions appropriately. We also encourage future work to investigate how algorithms can tailor health information recommendations in ways that mitigate availability bias, ensuring exposure to such content does not unintentionally heighten user anxiety or misperceptions.

## 7.5 Ethical Considerations

This study investigated how social media content about adult ADHD influences users' self-perception and explored the effective design of chatbot-based symptom checkers to mitigate availability bias in online self-diagnosis. In this health context, potential ethical issues should be carefully considered.

In terms of misleading information, we acknowledge that the social media content presented to participants during the study may cause misconception regarding ADHD. To address this, a debriefing session was provided at the end of the experiment to correct potential misconceptions.

In consideration of safety, while the online diagnostic tools employed were derived from validated medical scales, participants might blindly trust the produced results, raising ethical concerns. To prevent this, we avoided delivering diagnostic outcomes and instead offered general recommendations on seeking professional medical advice based on self-assessment scores. The debriefing session further emphasized that the diagnostic outcomes were for reference only and should not substitute professional guidance.

In online healthcare consultations, user privacy is a key consideration. In our study, we explicitly stated in the consent form that conversation data with the CSC would be used solely for research purposes and not shared with others without permission. Similarly, real-world CSCs should ensure confidentiality and anonymization of user data, particularly when handling sensitive medical information.

### 7.6 Limitations

We acknowledge several limitations in our study. Firstly, we chose adult ADHD as the case in this study as the prevalence of its symptoms allowed us to use the general public as a screening criterion for participants. Whether our findings generalize to other health conditions or disorders, especially when individuals already have presumed suspicions about certain disorder, needs to be explored in future studies. However, since cognitive biases have been observed to be a likely cause of other disorders' over-diagnosis such as Lyme disease from prior research [25], we have reason to believe that our results may generalize broadly.

Second, the study was short-term and conducted in a simulated environment. Participants conducted self-assessments immediately after reading social media posts, while individuals are likely to be exposed to health information intermittently over an extended period. Moreover, we retained original images in each social media post to enhance the realism of the simulated platform. This may introduce bias, as the visual content could influence participants' interpretation and engagement with the information. Additionally, users interact dynamically within social media communities, and their interaction data can generate further influence. Future work should attempt to explore how availability bias is triggered by users' interactions within real, dynamic online communities.

Then, Study 1 simulated three user scenarios: (1) no ADHD concerns and no exposure to ADHD-specific content, (2) ADHD concerns with exposure to factual ADHD content, and (3) ADHD concerns with exposure to exaggerated content. However, we assumed participants' ADHD baseline were uniform across three groups to attribute observed effects to the exposure conditions. According to our findings, users with strong ADHD concerns may experience greater availability bias than those without prior doubts about having ADHD, as the health information is more relevant to them. Thus, this limitation could lead to more conservative estimates of the effects of the neutral and exaggerated exposure conditions. Future work could extend this study by stratifying participants based on their initial ADHD concerns or self-assessment levels.

Lastly, our participants were skewed younger, were predominantly male, and had relatively higher educational levels compared to the general population. These biases may influence the generalizability of our findings. Previous research suggests that younger individuals and those with higher educational backgrounds are less susceptible to biased online information [70, 73], while gender differences show no significant impact [31]. Consequently, the observed effect of social media exposure on availability bias might be more pronounced in older and less educated populations, which warrants further investigation.

## 8 CONCLUSION

As health information-seeking behavior on social media increasingly shapes users' health perceptions, it becomes essential for individuals to resist availability bias triggered by social media exposure to make informed health decisions. In this work, we conducted two controlled experiments to investigate how social media exposure triggers availability bias and how to mitigate such bias through CSC designs. Through mixed-method approaches, we empirically found: 1) resonant social media content distorted users' self-perception during symptom assessment, leading to an overestimation of their symptoms; 2) CSCs with cognitive intervention strategies were effective in mitigating availability bias during the symptom self-assessment process. The contributions of our study lie in emphasizing the importance of considering potential cognitive biases in the design of online diagnostic tools, as well as providing preliminary insights into the effectiveness of integrating cognitive interventions within chatbots to reduce these biases.